\documentclass[a4paper,11pt]{article}

\usepackage{jcappub} 

\usepackage[T1]{fontenc} 
\usepackage{multirow,subcaption}

\title{\boldmath Research on electron and positron spectrum in the high-energy region based on the gluon condensation model}

\author[a]{Jin-tao Wu,}
\author[a]{Ming-jun Feng,}
\author[a,1]{Jian-hong Ruan\note{Corresponding author.}}

\affiliation[a]{Department of Physics, East China Normal University,\\Shanghai 201100, China}

\emailAdd{jin-taowu@qq.com}
\emailAdd{fengmingjun22@mails.ucas.ac.cn}
\emailAdd{jhruan@phy.ecnu.edu.cn}

\abstract{Electron(positron), proton and nuclei can be accelerated to very high energy by local supernova remnants (SNR). The famous excesses of electron and proton (nuclei) potentially come from such kind of local sources. Recently, the DAMPE experiment measured the electron spectrum (including both electrons and positrons) of cosmic rays with high-accuracy. It provides an opportunity to further explore the excess of electrons. According to the gluon condensation (GC) theory, once GC occurs, huge number of gluons condense at a critical momentum, and the production spectra of electron and proton showing typical GC characteristics. There are exact correlations between the electron and proton spectrum from a same GC process. It is possible to interpret the power-law break of cosmic rays in view of GC phenomenon, and predict one from another based on the relations between electron and proton spectrum. In this work, we point out the potential existence of a second excess in the electron spectrum, the characteristics of this excess is derived from experimental data of proton. We hope that the future DAMPE experiments will confirm the existence of this second excess and support the result of GC model.
	
 }

\begin{document}
\maketitle
\flushbottom

\section{Introduction}
\label{sec:intro}

In recent decades, much important developments have been made in space-based and ground-based cosmic rays (CRs) experiments. With the advent of these next-generation experiments, measurements of CRs are entering a high-precision era and revealing a range of new phenomena. In particular, in recent years, collaborations such as Alpha Magnetic Spectrometer (AMS-02) \cite{AMS:2019iwo,aguilar2021alpha}, Fermi Large Area Telescope (Fermi-LAT) \cite{Fermi-LAT:2017bpc}, DArk Matter Particle Explorer (DAMPE) \cite{DAMPE:2017fbg,DAMPE:2019gys}, and so on have provided more accurate experimental data of electron and positron (or $e^- + e^+$) and proton spectrum. Depending on the particle species, these data cover an energy range from about 1 GeV to several TeV. The CRs with energies above 1 TeV usually come from distances within 1 kpc from the solar system, highlighting the importance of nearby sources as detection targets. The energy spectra of CRs may be influenced by nearby sources such as pulsar wind nebulae (PWNe) \cite{DiMauro:2014iia,Hooper:2017gtd,Bykov:2019mis,Manconi:2020ipm,Evoli:2020ash,Bao:2020ila,Ding:2020wyk}, supernova remnants (SNRs) \cite{shen1970pulsars,Atoian:1995ux,Kobayashi:2003kp,di2014interpretation,Fang:2017tvj,Evoli:2020szd,Evoli:2021ugn}, and dark matter (DM) particle annihilation or decay \cite{Liu:2017rgs,Huang:2017egk,Coogan:2019uij,Liu:2019iik,Feng:2019rgm,Ge:2020tdh,Chen:2021yde}. A commonly accepted idea is that the detected spectrum should be a combination of contributions originating from background and local SNR sources. The energy spectra of both ($e^-+e^+$) and protons are precisely interpreted using this approach with appropriate nearby sources. The Payload for Antimatter Matter Exploration and Light-nuclei Astrophysics (PAMELA) \cite{PAMELA:2008gwm} discovered an excess of positron spectra above 20 GeV, a remarkable finding confirmed by the sharp drop in the positron spectrum observed by AMS-02 at an energy of 284 GeV \cite{AMS:2019rhg}. The DAMPE experiment found that the electron spectrum follows a power-law shape at energies up to approximately 0.9 TeV, followed by a spectral break-off at the same energy \cite{DAMPE:2017fbg}. Moreover, the electron spectrum exhibits a single power law below 0.9 TeV without any noticeable exotic features.

The AMS-02 experiment has also shown the significant excess in the nuclei (protons, He, C, O) spectrum at energies from $\sim$200 GeV \cite{AMS:2015tnn}. The hardening of the proton spectrum has been confirmed by the DAMPE experiment, which identified a significant excess of proton peaked at about 13.6 TeV \cite{DAMPE:2019gys}. This spectrum excess was previously established observationally by the ATIC \cite{Panov:2009iih} and NUCLEON \cite{Atkin:2018wsp}. The nucleon excess may also come from the same SNR \cite{Mertsch:2010fn,Bernard:2012wt,Serpico:2011wg,di2014interpretation,Fang:2018qco,Tang:2018wyr}.

The gluon condensation (GC) model, proposed by Zhu Wei et al., is obtained from the QCD kinetic evolution equation in condition of very high interaction energies \cite{Zhu:2008ur,Zhu:2016qif,Zhu:2017ydo,Zhu:2017bvp,Feng:2018wio}. GC is the state in which the gluons in proton (or nucleus) condense at a critical momentum, and the distribution of gluons appearing as delta-function like. When high-energy cosmic rays entering the atmosphere, the secondary particles such as electron-positron, pion, proton-antiproton, etc., are produced efficiently by proton-proton or proton-nucleus ($p - p(A)$) collisions. If GC happens, the producing electron and proton will be affected significantly and their energy spectra will show typical GC features. Specifically, since $p + p(A) \rightarrow p + \bar{p} + \pi^\pm + \pi^0 + \text{others}$, the excess of leptons observed in CRs may come from the secondary process $\pi^0 \rightarrow \gamma \rightarrow e^+ + e^-$ in the GC state. The proton-antiproton excess may also be derived from the same $p-p(A)$ collision. Therefore, an association between the spectra of electron and proton excesses and a correlation of GC energy thresholds between them is given in the literature \cite{Liu:2019gxw}. Based on the measurement of electron of DAMPE, we assume a second additional excess of electron with the position and magnitude estimated by the data of proton.

This paper is organized as follows. In section \ref{sec:2}, both the spatially-dependent propagation (SDP) model of CRs, the cosmic ray background SNR and its injection spectrum, local SNR and the GC model are introduced briefly. Section \ref{sec:3} presents computational results obtained by using the DRAGON2 \cite{Evoli:2016xgn,Evoli:2017vim} numerical package to solve the SDP function of the background CRs while considering synchrotron effects, inverse Compton scattering effects, and inverse Compton loss effects \cite{Strong:1998pw,Moskalenko:1997gh,Delahaye:2010ji}, and also solves the function of the local source and briefly analyzes them with experimental data. Finally, section \ref{sec:4} presents the discussion and summary.

\section{Model description}
\label{sec:2}

\subsection{Spatially-dependent propagation}
\label{subsec:2.1}

The CRs diffuses in the galactic halo via scattered magnetic waves and magnetohydrodynamic (MHD) turbulence after injection into interstellar space. The shape of the diffusive halo is usually approximated to be a cylinder, with its radial boundary equal to the Galactic radius $R=20 ~\text{kpc}$, while its half thickness $z_h$ needs to be fixed by the CRs data. Both CRs sources and interstellar medium (ISM) are mainly diffused in the Galactic disk, the width $z_s$ is roughly regarded as $200 ~\text{pc}$, which is much less than the halo’s thickness. In addition to the diffusion effect, the convection, reacceleration, and fragmentation processes exist in the transport of CRs particles by colliding with the ISM. Because of ionization and Coulomb scattering, CRs particles with low energies will further loss their energies. The transport equation is generally written as follows \cite{Maurin:2002ua,Strong:2007nh}:
	\begin{equation}
		\begin{aligned}
			\frac{\partial \psi}{\partial t} &= Q(\boldsymbol{r} , p)+\nabla \cdot\left(D_{x x} \nabla \psi-\boldsymbol{V}_{c} \psi\right)+\frac{\partial}{\partial p}\left[p^{2} D_{p p} \frac{\partial}{\partial p} \frac{\psi}{p^{2}}\right]  \\
			&-\frac{\partial}{\partial p}\left[\dot{p} \psi-\frac{p}{3}\left(\nabla \cdot \boldsymbol{V}_{c}\right) \psi\right]-\frac{\psi}{\tau_{f}}-\frac{\psi}{\tau_{r}}. 
		\end{aligned}
		\label{eq:2.1}
	\end{equation}
Here $\psi = \mathrm{d}n/\mathrm{d}p$ is the CRs density per particle momentum $p$ at position $\boldsymbol{r}$ and is associated with the flux as $\boldsymbol{J}=\psi c/4\pi $, $Q(\boldsymbol{r},p)$ is the source function, $D_{xx}$ and $D_{pp}$ are the diffusion coefficients in the coordinate and momentum space, $\boldsymbol{V}_c$ is the convection velocity, $\dot{p}$ is the energy loss rate, $\tau_{f}$ and $\tau_{r}$ are the fragmentation and radioactive decaying time scales. The momentum diffusion coefficient $D_{pp}$ is associated to $D_{xx}$ as $D_{pp} D_{xx} = \frac{4p^2 v_A^2}{3\delta (4-\delta^2) (4-\delta)}$, in which $v_A$ is the Alfvén velocity \cite{seo1994stochastic}.

The SDP model divides the entire diffusive region into two pieces. The Galactic disk and its surrounding areas are regarded as the inner halo (IH), and the vast regions outside IH are known as the outer halo (OH) \cite{Tomassetti:2012ga,Feng:2016loc}. The spatially-dependent diffusion coefficient can be parameterized as follows \cite{Zhang:2021ipu}:
	\begin{equation}
		\label{eq:2.2}
		D_{xx} (r,z,{R})=D_0 F(r,z) \beta^\eta \left( \frac{{R}}{{R}}_0 \right)^{\delta(r,z)}. 
	\end{equation}
Here $\beta = v/c$ is the velocity of CR particles, ${R}=pc/Ze$ is rigidity function, and the relationship between $F(r,z)$ and $\delta(r,z)$ and the source distribution function $F(r,z)$ can be described as \cite{Guo:2015csa,Liu:2018ujp}:
	\begin{equation}
		\label{eq:2.3}
		\begin{aligned}
			F(r, z)=\left\{\begin{array}{ll} \displaystyle
				g(r, z)+[1-g(r, z)]\left(\frac{z}{\xi {z_{h}}}\right)^{n}, & |z| \leq \xi {z_{h}} \\
				1, & |z|>\xi {z_{h}} ,
			\end{array}\right. 
		\end{aligned}
	\end{equation}
and
	\begin{equation}
		\label{eq:2.4}
		\begin{aligned}
			\delta(r, z)=\left\{\begin{array}{ll} \displaystyle
				g(r, z)+[\delta_0-g(r, z)]\left(\frac{z}{\xi {z_{h}}}\right)^{n}, & |z| \leq \xi {z_{h}} \\
				\delta_0, & |z|>\xi {z_{h}} .
			\end{array}\right. 
		\end{aligned}
	\end{equation}
Here $g(r,z) = N_m / [1+f(r,z)]$, the IH region indicated by the half-thickness of $\xi z_h$, which $\xi$ is determined by fitting the hardening of the proton spectrum, and the size of the OH region is $(1 - \xi) z_h$.

\subsection{Background supernova remnants}
\label{subsec:2.2}

The SNRs are generally considered as the most probable source for the acceleration of CRs. The charged particles are accelerated to a power-law distribution through diffusive shock acceleration. The distribution of the SNRs is approximated as axisymmetric, which is generally parameterized as:
	\begin{equation}
		\label{eq:2.5}
		f(r, z)=\left(\frac{r}{r_{\odot}}\right)^{\alpha} \exp \left[-\frac{\beta\left(r-r_{\odot}\right)}{r_{\odot}}\right] \exp \left(-\frac{|z|}{z_{s}}\right), 
	\end{equation}
where $r_\odot \equiv 8.5~\text{kpc}$ represents the distance from the solar system to the Galactic center. In this work, the parameters $\alpha$ and $\beta$ are taken as 1.09 and 3.87 respectively \cite{Green:2015isa}. The density of CRs sources descends exponentially along the vertical height from the Galactic plane, with $z_s = 100 ~ \text{pc}$. {The axisymmetric approximation is reasonable, given that the diffusion length of CRs typically exceeds the characteristic spacing between adjacent spiral arms by a significant margin. However, due to factors such as synchrotron radiation and inverse Compton scattering, the transport distance of energetic electrons is considerably shorter. Consequently, the aforementioned approximation may lose accuracy. Therefore, incorporating a more realistic representation of the spiral distribution is anticipated to exert a profound influence on the observed high-energy electron spectrum.}

{In this study, the spiral distribution of SNRs adhered to the model is established in \cite{Faucher-Giguere:2005dxp}. The Milky Way Galaxy comprises four major arms, and the path of the i-th arm is represented as a logarithmic curve: $\theta(r) = k^i \ln(r/r_0^i) + \theta_0^i$, where $r$ is the distance to the Galactic center. For the values of $k^i, r_0^i$ and $\theta_0^i$ for each arm, one can refer to \cite{Tian:2019bfy}.  Along each spiral arm, there is a spread in the normal direction which follows a Gaussian distribution, i.e.
	\begin{equation}\label{eq:f_i}
		f_{i}=\frac{1}{\sqrt{2\pi}\sigma_{i}}\exp\left[-\frac{(r-r_{i})^{2}}{2\sigma_{i}^{2}}\right],\quad i\in\left[1,2,3,4\right],
	\end{equation}
where $r_i$ is the inverse function of the $i$-th spiral arm's locus, and the standard deviation $\sigma_i$ is taken to be $0.07 r_i$. The number density of SNRs at various radii continues to adhere to the radial distribution in the axisymmetric case, i.e. eq. (\ref{eq:2.5}).} For the injection spectra of lepton and nuclei, we take a broken power-law form as given in paper \cite{Zhang:2021ipu},
	\begin{equation}
		\label{eq:2.6}
		{Q}={Q}_{0}\left\{\begin{array}{ll}
			\displaystyle \left(\frac{{R}_{\mathrm{br} 1}}{{R}_{0}}\right)^{\nu_{2}}\left(\frac{{R}}{{R}_{\mathrm{br} 1}}\right)^{\nu_{1}}, & {R} \leq {R}_{\mathrm{br} 1} \\ \displaystyle
			\left(\frac{{R}}{{R}_{0}}\right)^{\nu_{2}}, & {R}_{\mathrm{br} 1}<{R} \leq {R}_{\mathrm{br} 2} \\ \displaystyle
			\left(\frac{{R}_{\mathrm{br} 2}}{{R}_{0}}\right)^{\nu_{2}}\left(\frac{{R}}{{R}_{\mathrm{br} 2}}\right)^{\nu_{3}}, & {R}>{R}_{\mathrm{br} 2},
		\end{array}\right.
	\end{equation}
where ${Q}_0$, $\nu_{1,2,3}$, ${R}_0$ and ${R}_{\mathrm{br} 1,2}$ are the normalization, power index, reference rigidity, and broken rigidity respectively.

\subsection{Local source}
\label{subsec:2.3}

The local sources of CRs are found to be within $\sim$1 kpc around the solar system \cite{Yuan:2018rys,Luo:2021szz}. For the high-energy particles, the synchrotron in an interstellar magnetic field and inverse Compton scattering is far more important than fragmentation. So the eq.~(\ref{eq:2.1}) of leptons and nucleus can be simplified to \cite{Atoian:1995ux,Liu:2016gyv,Fornieri:2019ddi}:
	\begin{equation} 
		\label{eq:2.7}
		\frac{\partial \varphi}{\partial t}-D_{xx} \Delta \varphi + \frac{\partial}{\partial E}(\dot{E}\varphi) = Q_{\text{int}}(E,t) \delta(\boldsymbol{r} - \boldsymbol{r}^\prime). 
	\end{equation}
The energy loss rate $\dot{E}$ for leptons is approximated to $\dot{E} = -bE^2$ in the Thomson limit, and for the nuclei, this can be omitted reliably. The analytical solutions for leptons and protons are as follows \cite{Fornieri:2020kch}:
	\begin{equation} 
		\label{eq:2.8}
		\varphi(r,t,E) = \frac{Q(E)}{[4\pi D(E) t] ^{3/2}} \cdot \exp[-r^2/4D(E) t] . 
	\end{equation}
Here, the diffusion coefficient $D(E)$ only relate to energy, so eq.~(\ref{eq:2.2}) becomes
	\begin{equation} 
		\label{eq:2.9}
		D(E) = D_0 \beta^\eta \left( \frac{E}{E_0} \right)^{\delta_0}. 
	\end{equation}

\subsection{The GC Model}
\label{subsec:2.4}

The electrons and protons can be injected into the interstellar space from the local SNR, in which GC may happen.  If so, the following processes $p + p(A) \rightarrow \pi^{\pm} + \pi^0 +p + \bar{p} + (\text{others})$ and $\pi^0 \to e^+ + e^-$ will contribute to the corresponding  GC-characteristic energy spectrum. According to the theory of Zhu et al. \cite{Liu:2019gxw}, the positron and electron spectra take the following form
	\begin{equation} 
		\label{eq:2.10}
		\begin{aligned}
			Q_e(E) =\left \{ \begin{array}{ll}
				\displaystyle C_e (E_{\mathrm{GC}}) \left(\frac{E}{E_{\mathrm{GC}}}\right)^{-\beta_1} \left[\frac{1}{\beta_{2}}\left(\frac{E}{E_{\mathrm{GC}}}\right)^{-\beta_{2}}+ \left(\frac{1}{\beta_{3}}-\frac{1}{\beta_{2}}\right)\right], & \text { if } E \leq E_{\mathrm{GC}} \\
				\displaystyle \frac{C_e }{\beta_{3}} (E_{\mathrm{GC}}) \left(\frac{E}{E_{\mathrm{GC}}}\right)^{-\beta_{1} -\beta_{3}} , & \text { if } E>E_{\mathrm{GC}},
			\end{array}\right. 
		\end{aligned}
	\end{equation}
	and for protons 
	\begin{equation} 
		\label{eq:2.11}
		\begin{aligned}
			Q_p(E) =\left\{\begin{array}{ll}
				\displaystyle
				C_p \left(E_{\mathrm{GC}}\right)\left(\frac{E}{E_{\mathrm{GC}}}\right)^{-\beta_{2}}, & \text { if } E \leq E_{\mathrm{GC}} \\ \displaystyle
				C_p \left(E_{\mathrm{GC}}\right)\left(\frac{E}{E_{\mathrm{GC}}}\right)^{-\beta_{3} } ,  & \text { if } E >E_{\mathrm{GC}} .
			\end{array}\right. 
		\end{aligned}
	\end{equation}
Here the parameters $\beta_{1}$ and $\beta_{2}$ represent the propagating loss of electrons and gamma-rays or protons, respectively. $\beta_{3} = \beta_{2}+2\beta_{p}-1 $, where $\beta_{p}$ denote the acceleration mechanism of scattering protons. The coefficient $C_e (C_p)$ incorporates the kinematic factor into the flux dimension and the percentage of $\pi^0 \to 2\gamma$ and $\gamma + \gamma \to e^- + e^+$, $E_{\mathrm{GC}}$ is the GC-threshold energy for electron and proton, respectively.	
 
If there are two different GC sources, each one contributes to the electron and proton spectrum, we define the coefficient $C$ in eqs.~(\ref{eq:2.10}) and (\ref{eq:2.11}) as $C_e^{1}$, $C_p^{1}$ and  $C_e^{2}$, $C_p^{2}$ for each source respectively. According to work \cite{Liu:2019gxw}, there are definite relations betwwen the threshold energy of  $E_{\text{GC}(e)}$ and $E_{\text{GC}(p)}$
	\begin{equation}
		\frac{ E_{\text{GC}(e)} }{E_{\text{GC}(p)} } \simeq \frac{m_\pi}{2 m_p}, \label{eq:2.12}
	\end{equation}
and the coefficients  ${C_e^1}$ and ${C_p^1}$ of source 1,  ${C_e^2}$ and ${C_p^2}$ of source 2  have the relationship as
	\begin{equation}
		\label{eq:2.13}
		\frac{C_e^1}{C_p^1} = \frac{C_e^2}{C_p^2}.
	\end{equation}
In the following calculation, the above two equations play very important roles.

\begin{figure}[htbp]
	\centering 
	\subcaptionbox{\label{fig.1(a)}}{\includegraphics[width=0.5\textwidth]{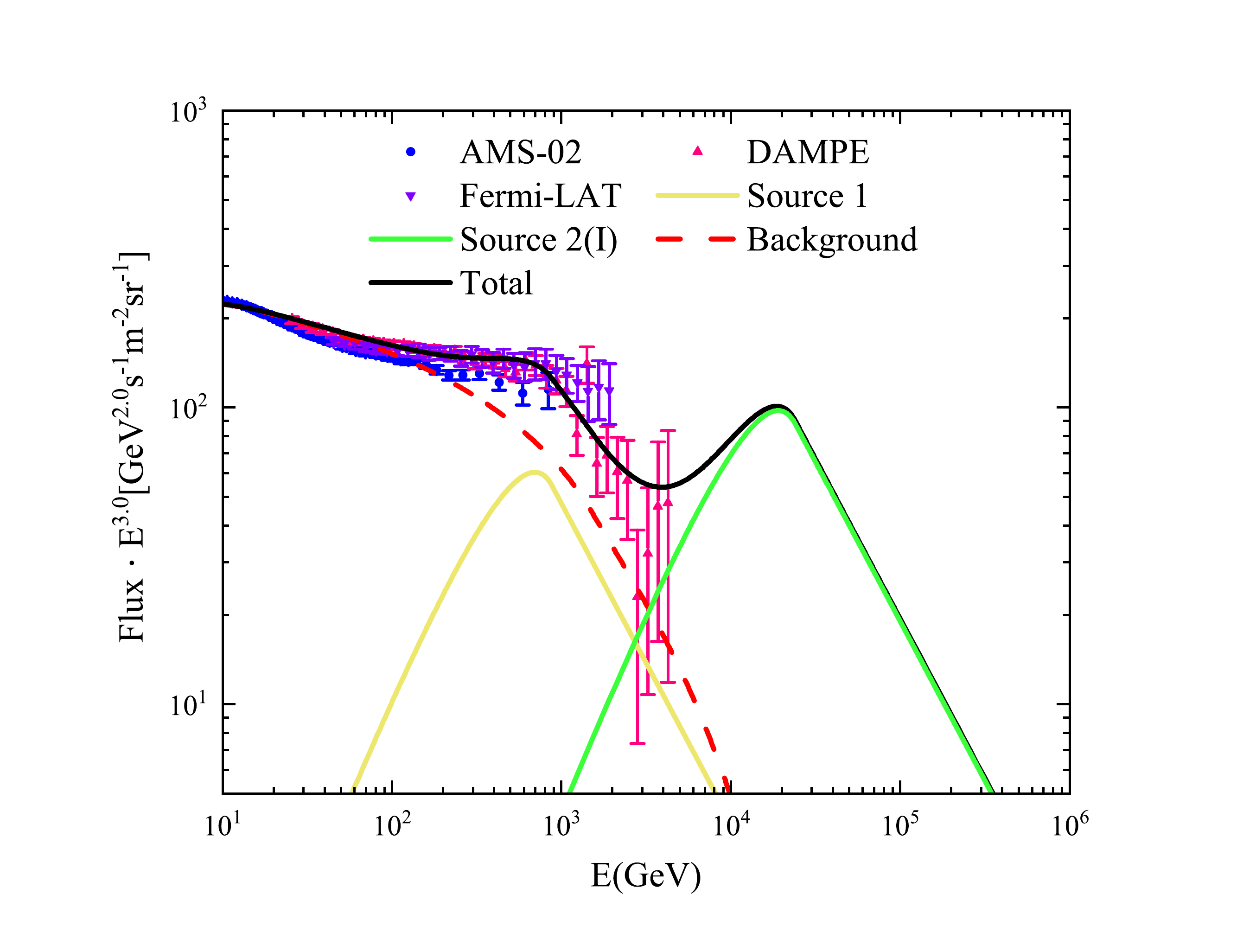}}\hfill
	\subcaptionbox{\label{fig.1(b)}}{\includegraphics[width=0.5\textwidth]{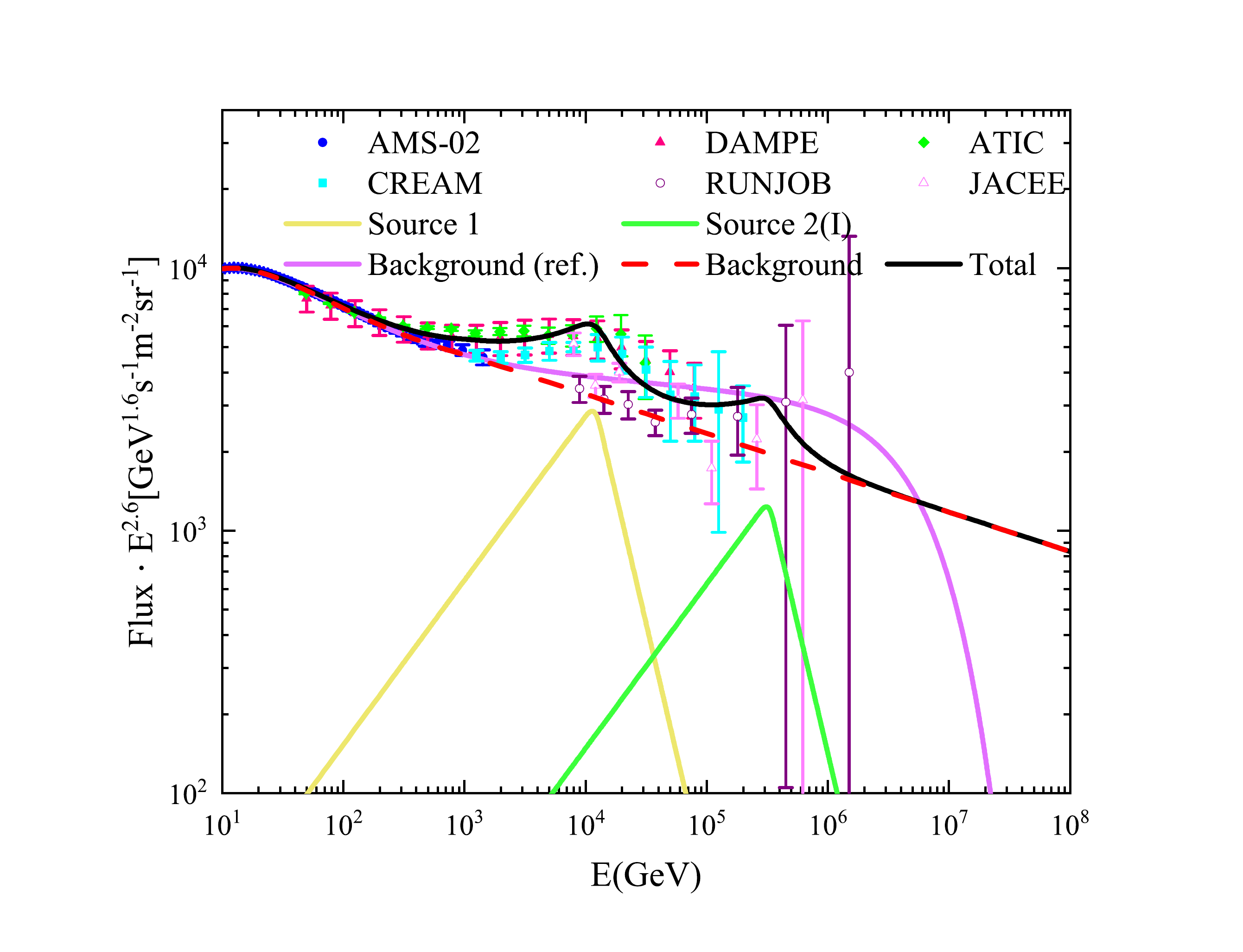}}
	
	\subcaptionbox{\label{fig.1(c)}}{\includegraphics[width=0.5\textwidth]{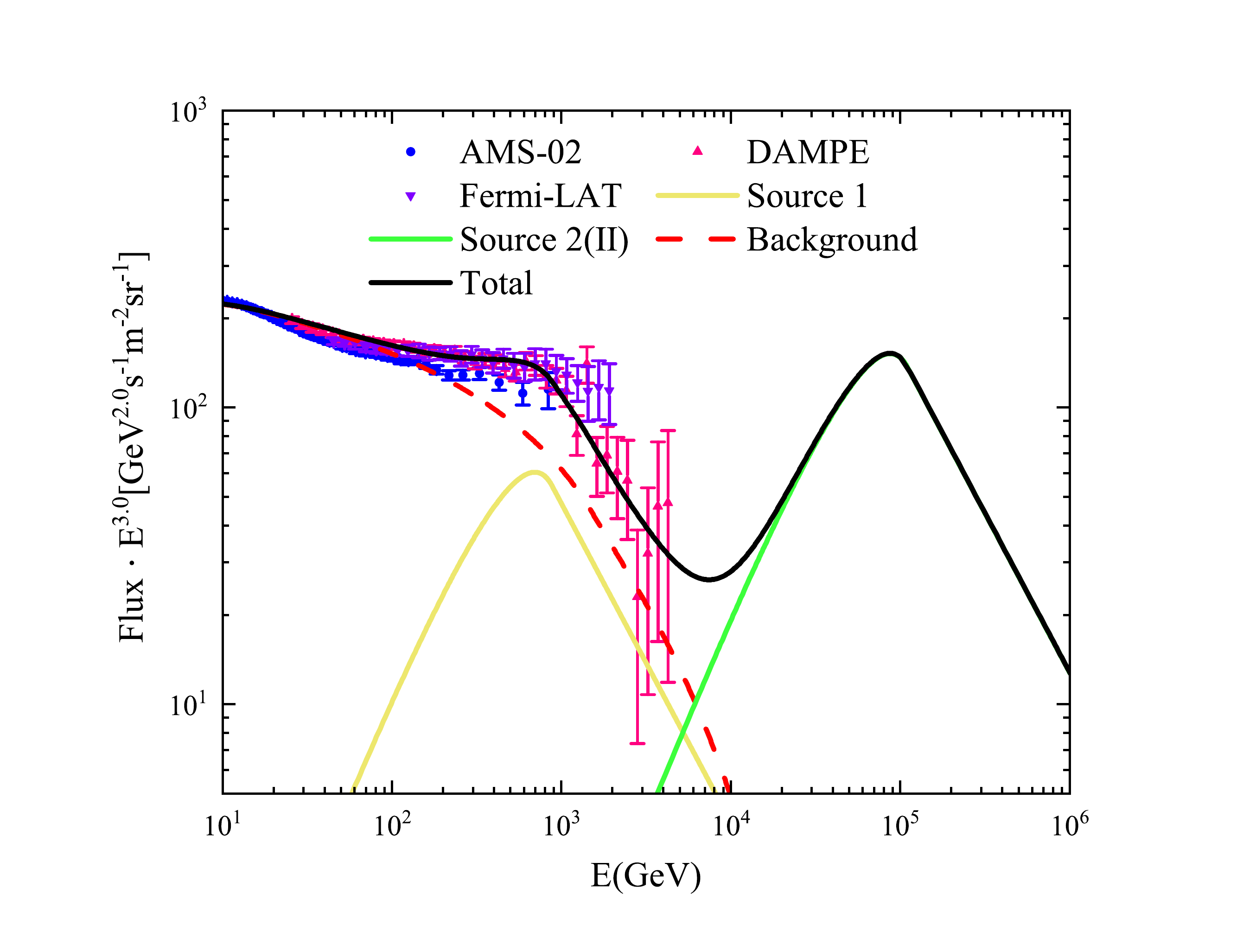}}\hfill
	\subcaptionbox{\label{fig.1(d)}}{\includegraphics[width=0.5\textwidth]{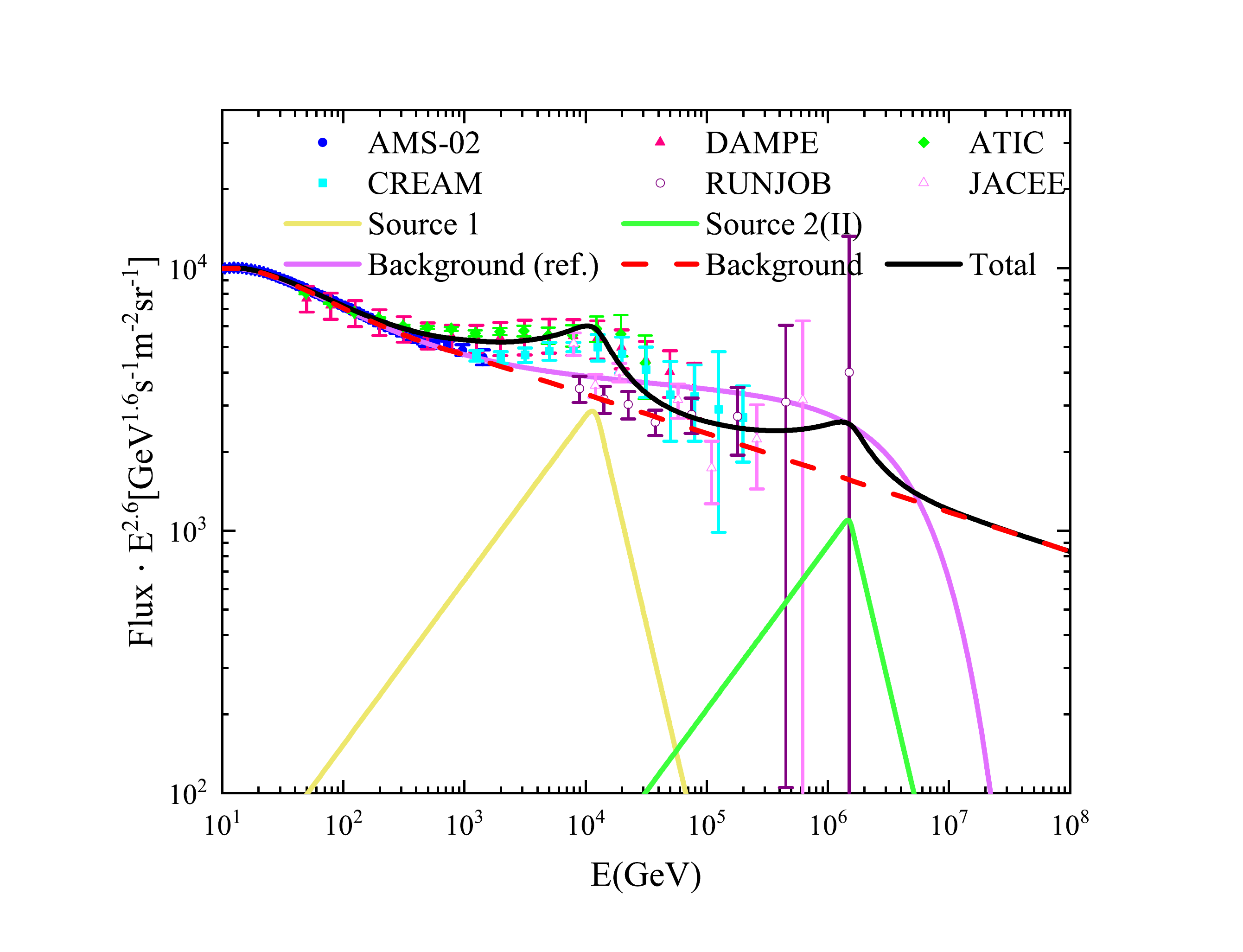}} 
	\caption{The right panel displays the calculated proton spectra, while the left panel shows the total spectra of positron and electron. Both panels feature red dashed lines representing the background spectrum, which was calculated by Dragon 2. The yellow and green lines represent the local source spectrum, while the black line represents the total spectra obtained by pulsing the background and local source spectra. The electron data points released by the AMS-02 (blue) \cite{AMS:2019iwo}, DMAPE (red) \cite{DAMPE:2017fbg}, and Fermi-LAT (purple) \cite{Fermi-LAT:2017bpc} experiments are shown in blue, red, and purple, respectively . Meanwhile, the proton data from the AMS-02 (blue) \cite{aguilar2021alpha}, DMAPE (red) \cite{DAMPE:2019gys}, ATIC (green) \cite{Panov:2009iih}, CREAM (azure) \cite{Yoon:2017qjx}, RUNJOB (purple) \cite{RUNJOB:2005mtb}, and JACEE (pink) \cite{Christ:1998zz} experiments were included.}
	\label{fig.1}
\end{figure}

\section{Prediction of the second excess of electron spectrum}
\label{sec:3}

The cosmic ray spectra of leptons and protons are typically considered to incorporate the contributions of local sources to the background. In general, the background propagation parameters are determined by protons, helium, B/C, and $^{10}$Be/$^{9}$Be ratios. In figure \ref{fig.1(a)} and figure \ref{fig.1(c)}, the  spectra of electrons are shown, while in figure \ref{fig.1(b)} and figure \ref{fig.1(d)} the spectra of protons are illustrated.

We assume that both the electron and proton spectra include contributions from two local GC sources, represented by the yellow and green solid lines, respectively. The first source (yellow, source 1) is probably from the Geminga SNR with an age of $3.24 \times 10^5 \text{~years}$ and locates at a distance of $0.33 \text{~kpc}$ from the solar system \cite{smith1994geminga}. The parameters for the second proton source (green), including its distance ($r$) and age ($t$), are inferred from the proton spectrum. Considering the larger uncertainty of proton data, we take two extreme possibilities for the second source, and denote them as source 2(I) and source 2(II).

The spectra of proton is depicted in figures \ref{fig.1(b)} and \ref{fig.1(d)}. Many models have been proposed to explain its characteristics. One type of models consider the spectra as one local source adding the background spectrum, which is represented by the solid pink lines in figures \ref{fig.1(b)} and \ref{fig.1(d)} \cite{Liu:2018fjy,Qiao:2019san}. This background line of proton (Background(ref.)) exhibits an abrupt drop around $2 \times 10^6 \text{~GeV}$, which is not a naturally expected power law decay. So, we prefer the red dashed line (Background) which is derived by fitting the AMS-02 experimental data and extending it. The observed spectra of proton can be fitted by using the Background and taking into account the contribution of two GC sources. However, the local source in high-energy region cannot be precisely determined because of large measurement uncertainties, and so, two extreme possibilities are considered and displayed as the green solid lines in figures \ref{fig.1(b)} (source 2(I)) and \ref{fig.1(d)} (source 2(II)). The GC effect of local sources contribute to the spectra excess of electron and proton. The spectra of electron (solid green line) in figures \ref{fig.1(a)} (source 2(I)) and \ref{fig.1(c)} (source 2(II)) are obtained by using equations eqs.~(\ref{eq:2.12}) and~(\ref{eq:2.13}) with the corresponding data of proton. To summarize, the analysis process is as follows:

(1). Determine the background spectra of electrons (represented by the red dashed line in figure \ref{fig.1(a)} and \ref{fig.1(c)}). We adopt the background propagation parameters for electron as reported in the work of \cite{Zhang:2021ipu,Tian:2019bfy}, $D_0 = 9.65 \times 10^{28} \mathrm{~cm}^2 \cdot \mathrm{s}^{-1}$ with a reference rigidity ${R}_0 = 4\mathrm{~GeV}$, $\delta_0 = 0.65$, $N_m = 0.24$, $\xi = 0.12$, $z_h = 5\mathrm{~kpc}$, and the Alfvén velocity is $v_A = 6 \mathrm{~km} \cdot \mathrm{s}^{-1}$, and remaining parameters are listed in table \ref{tab 1}. Using these parameters and the DRAGON2 package \cite{Evoli:2016xgn,Evoli:2017vim} to solve the transport equations, we get the background spectra of the electron.

(2). Determine the contribution of electrons from GC source 1 (represented by the solid yellow line in figure \ref{fig.1(a)} and \ref{fig.1(c)}). The location and age of source 1 are inferred from the experimental data. We assume that the GC source 1 may happen in Geminga SNR, the corresponding parameters are in table \ref{tab 2}. Using the DRAGON2 package, we calculate the contribution from GC source 1. For this source 1, the GC threshold energy is approximately $E_{\mathrm{GC}(e)}^{1} \simeq 880 \mathrm{~GeV}$.

(3). Determine the background spectra of protons (represented by the red dashed line in figures \ref{fig.1(b)} and \ref{fig.1(d)}). The purple solid line in both figures are one kind of background in works \cite{Liu:2018fjy,Qiao:2019san}, it's not naturally power law like and we suppose it should have complex combinations. In this work, we take the red dashed line as background which is the result by fitting the data of AMS02 and its extending.

\begin{table}[tbp]
	\centering
	\renewcommand\arraystretch{1.3}
	\begin{tabular}{|c|c|c|c|c|c|c|}
		\hline
		background  & $Q_0[\text{m}^{-2}\text{sr}^{-1}\text{s}^{-1}\text{GeV}^{-1}]$ & $\nu_1$ & ${R}_{br1} [\text{GeV}]$ & $\nu_2$ & ${R}_{br2} [\text{GeV}]$ & $\nu_3$  \\
		\hline
		electron & $1.18 \times 10^{-4}$ & 1.44 & 5.0 & 2.79 & 1500 & 3.8 \\
		\hline
	\end{tabular}  
	\caption{The injected spectra parameters of background for electron.}
	\label{tab 1}
\end{table}	

\begin{table}[tbp]
	\centering
	\renewcommand\arraystretch{1.3}
	\begin{tabular}{|c|c|c|c|c|c|c|}
		\hline
		GC source & source  & $C[\mathrm{GeV}^{-1}]$ & $E_\mathrm{GC}[\mathrm{TeV}]$ & $\beta_1$ & $\beta_2$ & $\beta_3$\\
		\hline
		\multirow{3}{*}{electron} & source 1  & $4.11 \times 10^{45}$ & 0.88 & 0.2 & 0.3 & 2.9\\
		\multirow{3}{*}{}      & {source 2(I)} & $1.33 \times 10^{46}$ & 24.0 & 0.2 & 0.3 & 2.9\\
		\multirow{3}{*}{}     & {source 2(II)} & $2.11 \times 10^{39}$ & 110 & 0.2 & 0.3 & 2.9 \\
		\hline
		\multirow{3}{*}{proton} & source 1  & $1.12 \times 10^{52}$ & 11.9  & - & 1.0 & 3.6\\
		\multirow{3}{*}{}       & {source 2(I)} & $3.71 \times 10^{47}$ & 324  & - & 1.0 & 3.6\\
		\multirow{3}{*}{}     & {source 2(II)} & $5.75 \times 10^{45}$ & 1500  & - & 1.0 & 3.6 \\
		\hline
	\end{tabular}  
	\caption{The injected spectra parameters of GC for electron and proton.}
	\label{tab 2}
\end{table}	

(4). Determine the proton spectrum from source 1 (represented by the solid yellow line in figure \ref{fig.1(b)} and \ref{fig.1(d)}). As we considering the first excess of electron spectrum from GC source 1, the source will also contribute an excess to the proton spectrum simultaneously. Using eqs. (\ref{eq:2.12}) and (\ref{eq:2.13}) in GC model, we determine the GC threshold energy as $ E_{\mathrm{GC}(p)}^{1} \simeq 11.9 \mathrm{~TeV}, C_p^1 \simeq 1.12 \times 10^{52}$. By using DRAGON2 package with the other parameters in table \ref{tab 2}, the contribution from source 1 is obtained.

(5). Determine the excess of electron and proton from source 2 (represented by the solid green line in figure \ref{fig.1}). Regarding electron data, the upper energy limit at present is approximately several TeV. So, in the higher energy region, we utilize proton data to estimate the contribution from source 2. For the proton spectrum, adding the contributions from the two sources to the background, it should be consistent with the data. While considering the larger uncertainties, we take two extremes as in \ref{fig.1(b)} (source 2(I)) and \ref{fig.1(d)} (source 2(II)).

For source 2(I), The GC threshold energy of electron $E^{\mathrm{2(I)}}_{\mathrm{GC}(e)} \simeq 24.0 \mathrm{~TeV}$, which is obtained by using the GC threshold energy of proton $E^{\mathrm{2(I)}}_{\mathrm{GC}(p)} \simeq 324.0 \mathrm{~TeV}$ with eq.~(\ref{eq:2.12}) \cite{Zhu:2017bvp}. Equation (\ref{eq:2.13}) is used to determine the relationship between $C_e$ and $C_p$ for sources 1 and 2. From experimental data, we determine $C_{e}^{1} \simeq 4.11 \times 10^{45}$ for electron (source 1) and $C_{p}^{1} \simeq 1.12 \times 10^{52}$ for proton (source 1). Thus, only $C_e$ or $C_p$ of source 2 needs to be confirmed. Comparing with data of proton, we found $C_{p}^{\mathrm{2(I)}} \simeq 3.71 \times 10^{47}$ at  GC threshold enery of $324.0 \mathrm{~TeV}$. Substituting this value, along with $C_{e}^{1}$ and $C_{p}^{1}$ into eq.~(\ref{eq:2.13}) yields $C_{e}^{\mathrm{2(I)}} \simeq 1.33 \times 10^{46}$, with a corresponding GC threshold energy of $24.0 \mathrm{~TeV}$. Using the remaining parameters listed in table \ref{tab 2}, the electron and proton spectra of source 2(I) in figures \ref{fig.1(a)} and \ref{fig.1(b)} (solid green line) are obtained. 

For source 2(II), considering the maximum energy of RUNJOB \cite{RUNJOB:2005mtb}, the maximum GC threshold energy of proton source 2 is assumed to be $E^{\mathrm{2(II)}}_{\mathrm{GC}(p)} \simeq 1500 \mathrm{~TeV}$, accordingly, for electron of source 2, $E^{\mathrm{2(II)}}_{\mathrm{GC}(e)} \simeq 110 \mathrm{~TeV}$. The coefficient for proton source 2 with GC threshold energy of $1500 \mathrm{~TeV}$ is $C_{p}^{\mathrm{2(II)}} \simeq 5.75 \times 10^{45}$, which in turn yields the $C_e$ for electron source 2 with threshold energy of $110 \mathrm{~TeV}$ , which is $C_{e}^{\mathrm{2(II)}} \simeq 2.11 \times 10^{39}$. Along with other parameters in table \ref{tab 2}, the electron and proton spectra of source 2 in figures \ref{fig.1(c)} and \ref{fig.1(d)} (solid green line) are obtained.

\section{Discussion and summary}
\label{sec:4}
How about our model comparing with some other models?

(1) In the above calculation, we only considered one set of background model propagation parameters. However, within the measurement uncertainties, there are some other diffusion model parameters that can fit the data reasonably well, for example \cite{Yuan:2018vgk,Tang:2021ozn}. 

\begin{figure}[htbp]
	\centering 
	\subcaptionbox{\label{fig.2(a)}}{\includegraphics[width=0.5\textwidth]{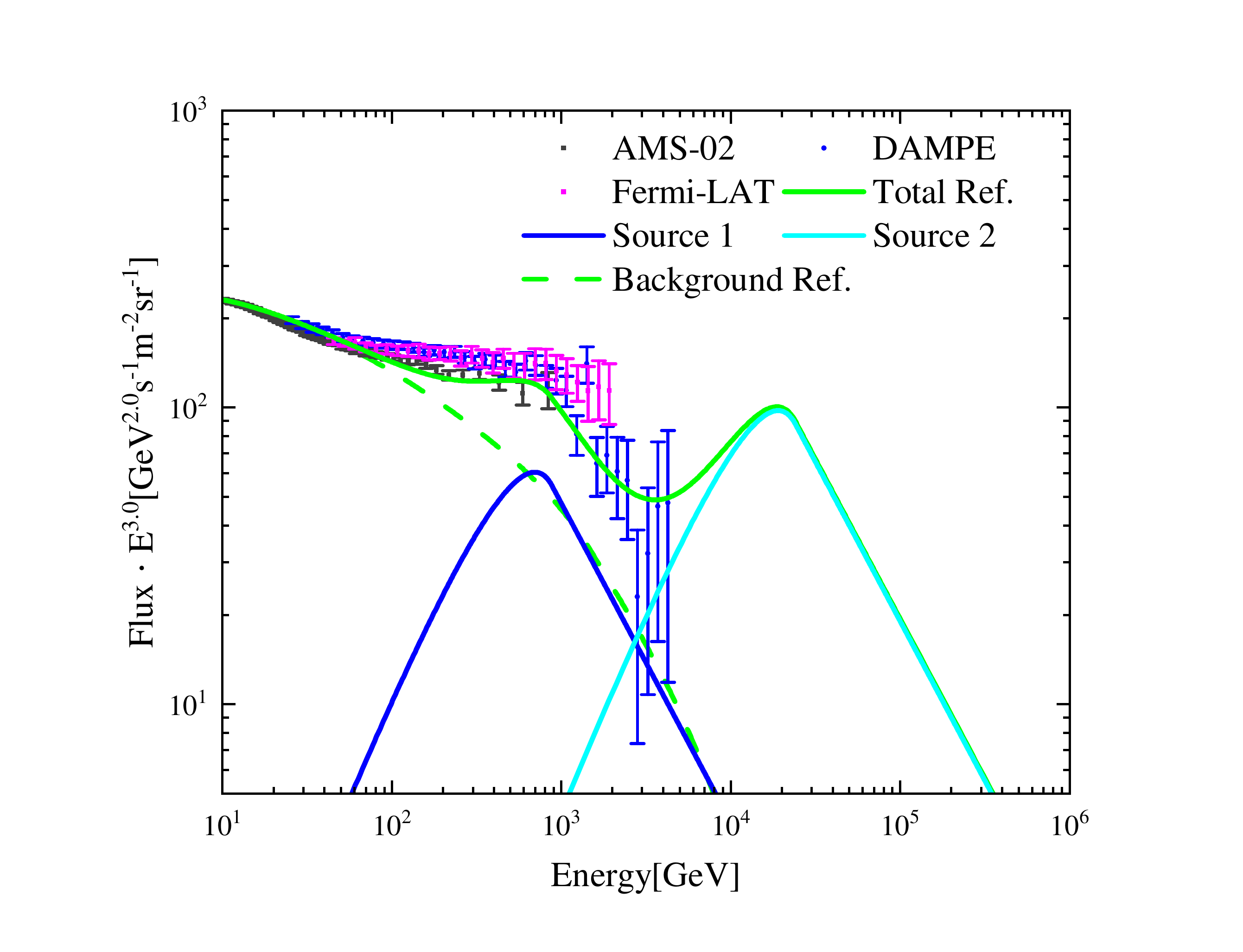}}\hfill
	\subcaptionbox{\label{fig.2(b)}}{\includegraphics[width=0.5\textwidth]{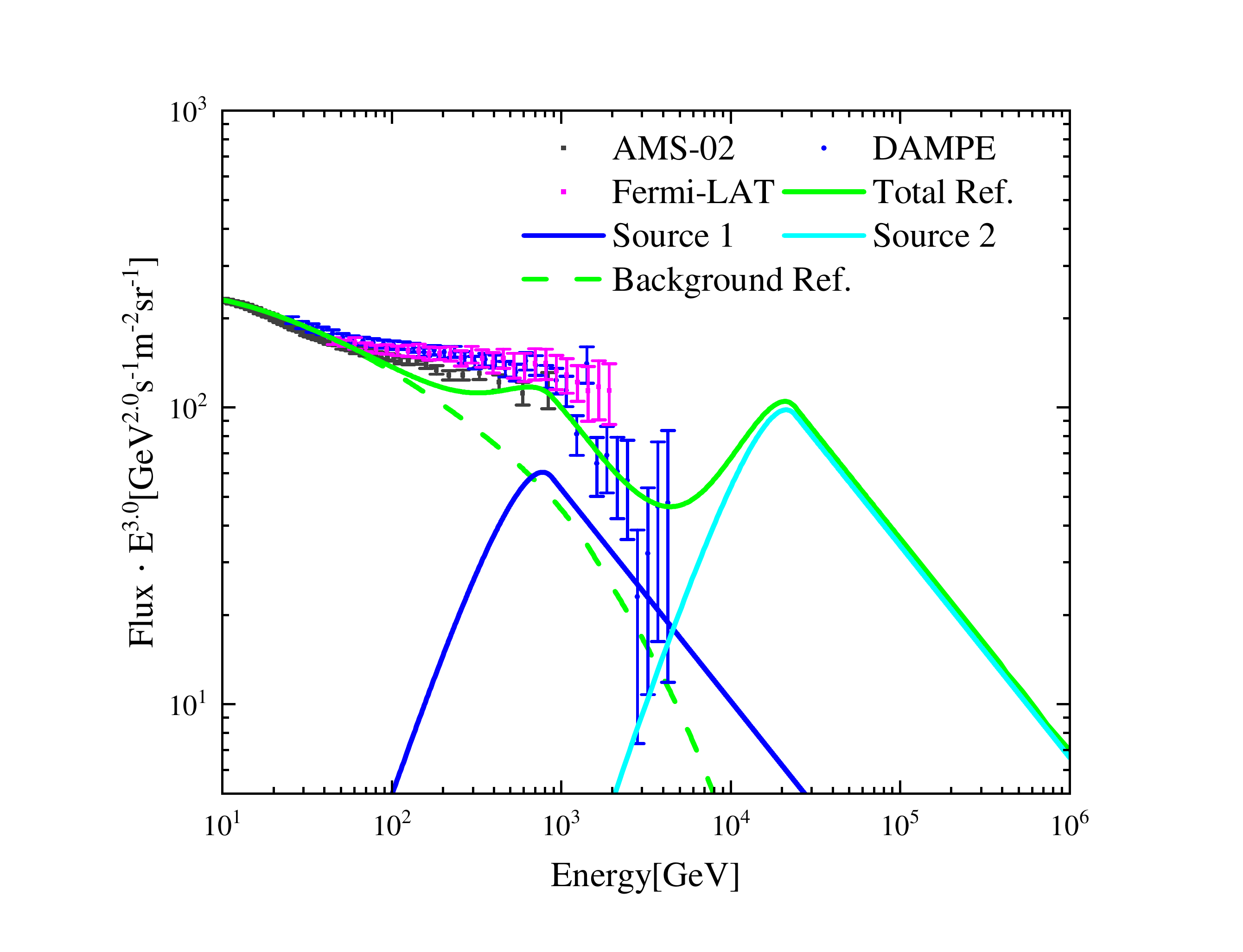}}
	\caption{Comparative diagrams of different propagation parameters. In figure \ref{fig.2(a)}, the green dashed line corresponds to the electron background spectra obtained using the propagation parameters from the article \cite{Yuan:2018vgk,Tang:2021ozn}, and the solid green line represents the total electron spectra. Figure \ref{fig.2(b)} depicts the electron spectra of local sources using the reference propagation parameter data.}
	\label{fig.2}
\end{figure}

For comparison, we take another set of propagation parameters as in \cite{Yuan:2018vgk,Tang:2021ozn}, where the halo size $z_h = 6.11$, the diffusion constant $D_0 = 6.46 \times 10^{28} \mathrm{~cm}^2 \cdot \mathrm{s}^{-1}$, the rigidity-dependence slope of the diffusion coefficient $\delta_0 = 0.41$, and the Alfvén velocity $v_A = 29.4 \mathrm{~km} \cdot \mathrm{s}^{-1}$. With these parameters and utilizing the DRAGON2 package, the cosmic ray propagation background was computed, which is shown as the dashed green line "Background Ref" in figure \ref{fig.2(a)}, and the total line which includes the two GC sources is shown as  the green solid line. Comparing figure \ref{fig.2(a)} with figure \ref{fig.1(a)}, it is evident that the background spectra of electrons (red dashed line in \ref{fig.1(a)}  ), as determined in this study, exceeds that of the reference background spectrum (green dashed line in \ref{fig.2(a)} ). The total electron spectra depicted in this work(solid black line in \ref{fig.2(a)}) aligns more favorably with the data from DAMPE, while the total reference spectrum (solid green line in \ref{fig.2(a)}) exhibits better conformity with the data from AMS-02. However, when the propagation parameters for the local sources are also adjusted to match the reference propagation parameters as in \cite{Yuan:2018vgk,Tang:2021ozn}, and the electron spectrum of the two local sources, as shown in figure \ref{fig.2(b)}, are obtained, the total electron spectra do not fit the experimental data well. So, it seems that the propagation parameters for the local sources in \cite{Yuan:2018vgk,Tang:2021ozn} are not suitable for our GC model.
 
 (2) In this work, we propose a second excess of electron spectrum according to GC model. There are also some other models which discuss about  a second peak \cite{Tang:2021ozn, Yue:2019sxt}. 
 
 The work by \cite{Yue:2019sxt} introduces a model involving multiple populations of CRs sources. In this model, the observed spectral bumps in protons around 10 TeV are attributed to the cutoff of population I, characterized by a cutoff rigidity of 60 TeV. Subsequently, the spectrum exhibits increased hardness for rigidities exceeding 100 TeV, which is attributed to contributions from population II. Population II is characterized by a cutoff rigidity of approximately 4 PeV, corresponding to the knee observed in the all-particle spectrum. In work by \cite{Tang:2021ozn}, they propose an explanation for the excess in the primary spectrum of electron using a single local SNR source and consider a second excess in the spectrum. They suggest that CR electrons with energies above 1 TeV might primarily originate from Vela SNR. However, in both of these works \cite{Tang:2021ozn, Yue:2019sxt}, they didn't relate the second proton excess with the second electron excess, they just discussed one of them. 
   
  While in our GC model, the excess of electron is closely related to the excess of protons. Within a GC process, described as $p + p(A) \rightarrow \pi^\pm + \pi^0 + p + \bar{p} + \text{(others)}$ and $\pi^0 \rightarrow e^+ + e^-$, the  injection spectra for electrons and protons are shown as eqs.~(\ref{eq:2.10}) and (\ref{eq:2.11}) respectively. The excesses of electrons and protons originating from the same GC source shows definite relationship between their GC energy thresholds as eq.~(\ref{eq:2.12}). And for the excesses from two different GC sources, eq.~(\ref{eq:2.13}) establishes the relationship between their $C_e$ and $C_p$ for electron and proton of sources 1 and 2. So we can estimate one from another. With the electrons and protons data from AMS-02 and DAMPE measurements, we assume that the first excesses are from GC source 1. In higher energy region, the data of proton shown another excess, if it is from a GC source 2, there should be a second excess in the spectra of electron too.  Because of the large uncertainties of the data (proton) in high energy region, the GC energy thresholds of the second excess of electron and proton cannot be accurately determined, so we consider two extreme possibilities in this work.

\begin{figure}[htbp]
	\centering 
	{\includegraphics[width=0.5\textwidth]{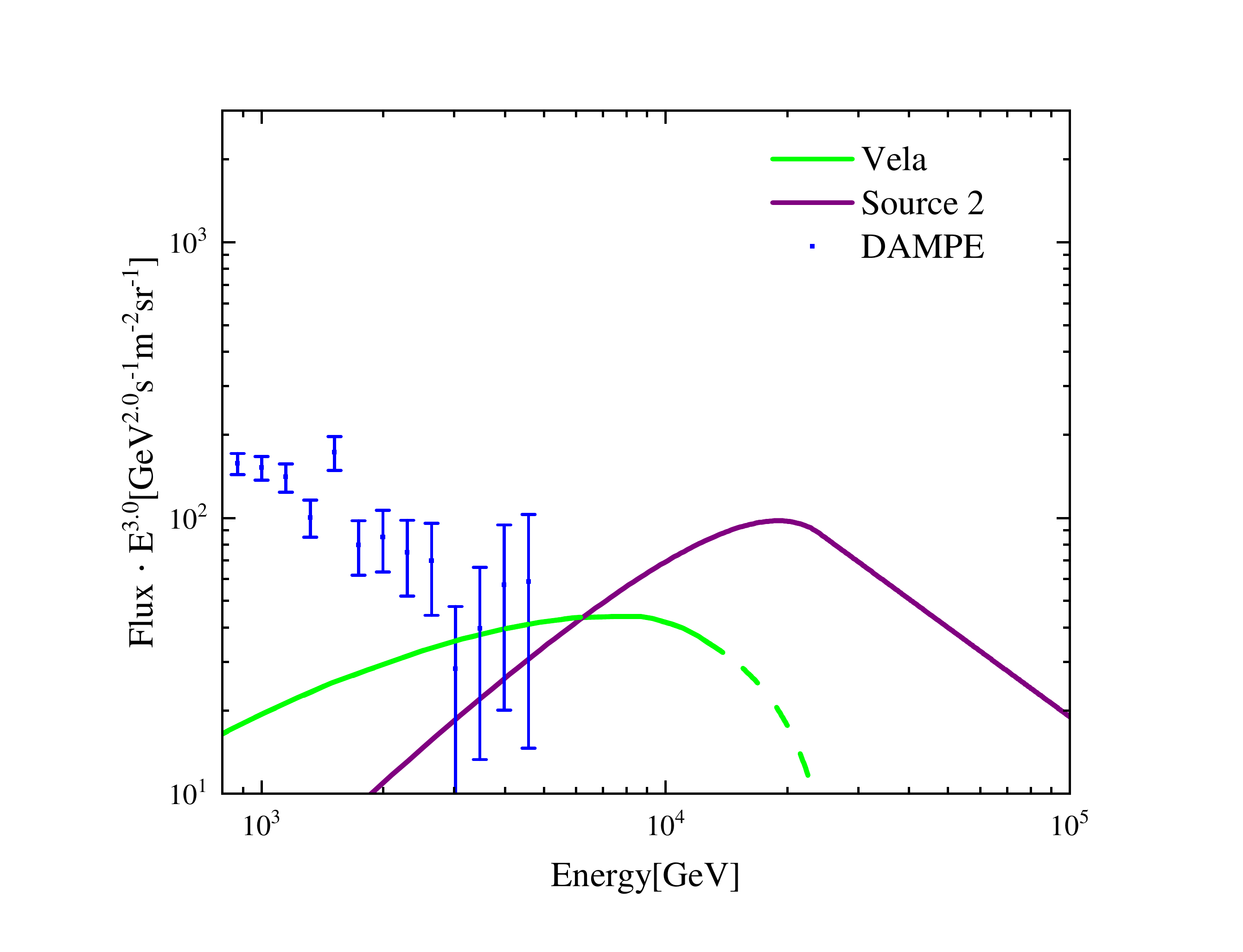}}
	\caption{Comparison between Vela and Source 2 in this paper. The solid green line represents the electron spectra contributed by Vela as presented in \cite{Tang:2021ozn}, while the dashed green line represents the curve obtained by interpolating and extrapolating the data from the mentioned reference using Origin software. The solid purple line represents the electron spectra contributed by Source 2 in this paper.}
	\label{fig.3}
\end{figure}

In summary, according to the proton measurements, we assume that the spectra of electron exist a second excess. The energy thresholds for these excesses in both electron and proton spectra are determined through the GC model and measurements. We find that the observed first excess in the electron spectrum is near $E_{\mathrm{GC}(e)}^{1} \simeq 880 \mathrm{~GeV}$ and that of proton $E_{\mathrm{GC}(p)}^{1} \simeq 11.9 \mathrm{~TeV}$ which may originate from the Geminga SNR. Additionally, We further hypothesize that the minimum possible second excess in the electron spectrum is near $E_{\mathrm{GC}(e)}^{\mathrm{2(I)}} \simeq 24.0 \mathrm{~TeV}$ and in proton spectrum is about $E_{\mathrm{GC}(p)}^{\mathrm{2(I)}} \simeq 324 \mathrm{~TeV}$, while the maximum of the possible second excess in electron spectrum is near $E_{\mathrm{GC}(e)}^{\mathrm{2(II)}} \simeq 110 \mathrm{~TeV}$ and in proton spectrum is about $E_{\mathrm{GC}(p)}^{\mathrm{2(II)}} \simeq 1500 \mathrm{~TeV}$. Presently, measurements of electrons above $40 \mathrm{~TeV}$ are still in large uncertainties. In Figure \ref{fig.3}, we compare our suggested second excess (purple solid line) with the model proposed by \cite{Tang:2021ozn} (green solid line, the dashed green part is our extension of the solid line), one can see that the difference is very obvious. Our spectrum of Source 2 is considerably higher, with a significantly larger peak energy. We anticipate that forthcoming precise measurements will validate the models.

\acknowledgments

This work is supported by the National Natural Science Foundation of China (No.11851303). 



\bibliographystyle{JHEP} 
\bibliography{ref} 



\end{document}